# Enterprise Architecture in Healthcare Systems: A systematic literature review

*Silvano Herculano da Luz Júnior[a], Francisco Ícaro Cipriano Silva[a], Gustavo Sousa Galisa Albuquerque[a], Francisco Petrônio Alencar de Medeiros[a], Heremita Brasileiro Lira[a]*

**Abstract**

*Purpose*: To present an overview of the state of the art of the application of Enterprise Architecture (EA) in healthcare systems to reveal which methodologies and tools are most used and which were the criteria for their choice. Identify the main positive impacts, challenges, and critical success factors described by the articles' authors and list the main publication channels and authors who have published the most on the topic.

*Methods*: A systematic literature review protocol (SLR) that included four phases and nine specific research questions. The scientific databases used were Science Direct, IEEE, Hubmed, and Scopus. As inclusion criteria, the study should be complete and answer at least one of the research questions. Researchers assessed the level of disagreement during the studies' evaluations using Cohen's Kappa. Five researchers independently read the included studies, performed the analysis, and compared the results. Conflicts were resolved by consensus through a disagreement meeting.

*Results*: The SLR protocol returned 280 primary studies after the first step of the Data Selection and a consolidated inclusion of 46 articles after the second step. It was possible to identify TOGAF as one of the most used frameworks and Archimate as one of the most implemented models in constructing EA in health systems. Criteria such as flexibility, interoperability, and being open source are the most described by the authors to choose EA methodologies and tools. It was possible to understand that the organizational complexity of healthcare environments and the difficulty of integrating data from various sources are among the main problems and challenges encountered in implementing EA in healthcare systems. In our analysis, the results of the categorization of the main positive impacts described by the authors highlight that the EA describes and categorizes the architecture and operation of business processes, the organizational structure, and the data to facilitate the acquisition of information and which brings benefits to the management changes and to improve quality processes, among other benefits mentioned. It was also possible to find in two articles the critical success factors for implementing EA in health systems, and we have listed 15 publication channels and nine authors who have published the most on the topic.

*Conclusions*: Although there are many articles on implementing EA in health systems and the benefits achieved by these works are also quite evident, most publications lack detailed and accurate information about the application environment or other data that could be relevant for the dissemination of good practices and the success achieved. Their inaccuracy and lack of detail often make the extraction of data a challenging job for conducting research. On the other hand, studies that have provided details about the health environment's nature offered a significant scientific basis for other organizations that seek methodologies, methods, and tools to assist their management and governance. These studies can become an essential empirical basis in selecting a set of good practices and making it possible to carry out new relevant studies.

## 1. Introduction

The Enterprise Architecture (EA) had its first studies conducted from the '80s by John A. Zachman, defining information systems architecture by creating a descriptive framework from disciplines that were wholly independent of information systems. By analogy specifies information systems architecture based on a neutral, objective framework (Zachman, 1987). However, at this time, there was little interest in the idea of Enterprise Engineering or Enterprise Modeling. The use of formalisms and models was generally limited to some aspects of application development within the Information Systems community (Zachman, 2016).

As technology advances, there is an increasing immersion of Information Technology (IT) in the provision of healthcare services. Complex systems distributed in a way that speeds up and accrete value to the hospital business, with faster responses that assist healthcare professionals in decision making that directly or indirectly impact patients. EA seeks to reflect the complexity of modern IT systems, which comprise hundreds of components, organized in different layers, with many relationships among them (Yoo et al., 2010).

[a] Federal Institute of Paraíba, Informatics Department, Avenue Primeiro de Maio, 720, Jaguaribe, João Pessoa - Paraíba, Brazil.

EA distinguishes the current and desirable future states of an organization's processes, capabilities, application systems, data, and IT infrastructure and provides a roadmap for achieving this target from the current state (Ross et al., 2006; Tamm et al., 2011). Gartner (2020) defines Enterprise architecture (EA) as a discipline for holistically leading enterprise responses to disruptive forces by identifying and analyzing the change's execution toward desired business vision and outcomes. EA delivers value by presenting business and IT leaders with signature-ready recommendations for adjusting policies and projects to achieve targeted business outcomes that capitalize on relevant business disruptions. EA enables business-driven and IT-driven change opportunities. Each of these organizational change processes leads to project benefits, resulting in organizational benefits (Shanks et al., 2018).

This research consists of a deep and broad systematic review of the literature to select studies that demonstrate current practices of Enterprise Architecture in Healthcare Systems. The search string for this work selected studies carried out between 2015 and 2019, which had practical applications of EA in several branches of Healthcare Systems, which selected 46 studies. The research was carried out by two teams of researchers, and the Kappa method assessed the level of disagreement during the evaluation of the studies by the researchers. The protocol in this research complies with the guidelines and procedures of Kitchenham (2007) for conducting Systematic Literature Reviews in Software Engineering and complemented by the approach of Dybå & Dingsøyr (2008).

There are two main contributions of this work. First, it gathers studies that provide relevant information on the state of the art of EA in Healthcare Systems, explaining the current primary practices, so that researchers and professionals in this area can take advantage of this information, intensify good practices and improve governance and healthcare management. Second, it seeks to intensify the performance of systematic reviews of the literature addressing EA in healthcare systems, since, in our research scope, we did not find reviews that deal with this topic specifically.

This paper is structured as follows. Section 2 presents a theoretical foundation with a search of the main concepts of EA in the literature, where it explains some most used methodologies, frameworks, and tools. Section 3 addresses the research methodology used in the systematic literature review process. Section 4 discusses the evaluation of the synthesis results, gathering the artifacts that answer the research questions analyzed in this work. Finally, Section 5 shows some findings and contributions to research and practice, as well as a discussion of directions for future research.

## 2. Theoretical foundation

Enterprise Architecture is a management and technology practice that is devoted to improving enterprises' performance. It enables them to see themselves in terms of a holistic and integrated view of their strategic direction, business practices, information flows, and technology resources (Bernard, 2012). EA includes details about an organization's processes, capabilities, data, application systems, and IT infrastructure using a variety of standardized representation techniques (Kaisler et al., 2005; Lankhorst, 2013). An enterprise-wide architecture should serve as an authoritative reference, source of standards for processes/resources, and provider of designs for future operating states. Moreover, the best practices are very resource-intensive, and the scope is not all-inclusive because of the costs of implementation and maintenance methods. The organization faces the challenge of deciding which to adopt, how to do it, and what overlaps, contradictions, and gaps from the resulting collection (Bernard, 2012).

Processes based on ontological structures are predictable and produce repetitive results; on the other hand, those not based rely exclusively on the expertise of their practitioners (Zachman, 2016). EA's analysis is not limited to IT systems, but also covers the relationship and support of business entities. Thus, EA research tends to focus on the "strategic" implications of EA's efforts in the Mission, Vision, Strategy, Objectives, Actions, and Operations of the analyzed business systems (Aier, 2014; Boh & Yellin, 2007; Ross et al., 2006).

When dealing with healthcare environments, as technology advances, there is an increasing immersion of Information Technology (IT) in the provision of healthcare services. These complex systems are distributed in a way that speeds up and accretes value to the hospital business, with faster responses that assist healthcare professionals in decision making that directly or indirectly impact patients. EA seeks to reflect the complexity of modern IT systems, which comprise hundreds of components, organized in different layers, with many relationships among them (Yoo et al., 2010).

There are currently several EA frameworks that can assist organizations in their management and governance processes. Each of them has characteristics that can suit certain types of organizations and their specific needs. For example, in a comparative study among several EA frameworks, carried out by Haghighathoseini et al. (2018), the TOGAF (The Open Group Architecture Framework) is the most appropriate for hospitals. The TOGAF was first developed in 1995 and is based on the DoDAF (Department of Defense Architecture Framework). This framework provides a structure for final product descriptions and a set of guidelines and rules for the consistency of these products, ensuring a common basis for integration and comparison of systems and families of systems, interoperating their architectures (Defense, 2010).

The ArchiMate® Specification, a Standard of The Open Group, is an open and independent modeling language for Enterprise Architecture that is supported by different tool vendors and consulting firms. It defines a common language for describing the construction and operation of business processes, organizational structures, information flows, IT systems, and technical infrastructure. This insight helps stakeholders to design, assess, and communicate the consequences of decisions and changes within and between these business domains (Group, 2020)

ISO/IEC 42010:2007 defines "architecture" as: "The fundamental concepts or properties of a system in its environment embodied in its elements, relationships, and in the principles of its design and evolution". TOGAF embraces but does not strictly adhere to ISO/IEC 42010:2007 terminology and describes a twofold definition of "architecture": depending on the context: 1. A detailed plan or a formal description of a system, at the component level, to guide its implementation; 2. The structure of the components, their interrelations, principles, and guidelines govern their design and evolution over time.

The AIDAF (Adaptive Integrated Digital Architecture Framework) is a model of an EA framework integrating an adaptive EA cycle for different business units. It involves the Architecture Board performing architecture reviews and enabling the alignment between IT architecture strategy and each solution architecture in Information System or IT projects, including digital IT solutions (Masuda et al., 2017).

However, given the various EA frameworks cited in the literature, this systematic review selects studies between 2015 and 2019 that implement these components in healthcare environments, gathering useful information for researchers and professionals in the field.

### *2.1. Related works*

During this systematic literature review process, six papers address relevant issues that contribute to the gathering of knowledge on the EA state of the art. Table 1 lists these reviews found.

**Table 1**
Related studies

| **Study** | **Reference** |
|---|---|
| A Systematic Review of Enterprise Architecture Assessment Models | (Bakar et al., 2015) |
| Systematic Literature Review on Enterprise Architecture in the Public Sector | (Dang & Pekkola, 2017) |
| A systematic literature review: Critical Success Factors to Implement Enterprise Architecture | (Ansyori et al., 2018) |
| A Systematic Review of Business-IT Alignment Research with Enterprise Architecture | (Zhang et al., 2018) |
| Past, current and future trends in enterprise architecture - A view beyond the horizon | (Gampfer et al., 2018) |
| The value of and myths about enterprise architecture | (Gong and Janssen, 2019) |

In Ansyori et al. (2018), the review addresses the state of the art of EA implementation in the public sector, focusing on critical success factors, which differ from the main objective of this research, which focuses on healthcare systems, public or private. Zhang et al. (2018) performed a Systematic Literature Review about BITA (Business-IT Alignment) using EA. They answered six questions through the 5W1H method to understand BITA from the perspective of EA. In Bakar et al. (2015), the review's main objective is to identify and categorize the existing EA assessment models according to the EA phases and analyze the model's limitations. Dang & Pekkola (2017) sought, through a systematic literature review, to identify the main topics and research methods in studies focused on public sector EA, where 71 articles identified in the last 15 years were analyzed. The authors' analysis shows that the development point of view, the case studies in developed countries, and the local configurations seem to form the primary research of EA in the public sector.

In the article by Gampfer et al. (2018), on the other hand, the EA is discussed more holistically, aiming to provide an overview of its evolution over 30 years, as well as identify the current status of the EA and detect trends that can most impact the Enterprise Architecture in the upcoming years. For Gong & Janssen (2019), the review seeks to identify the myths that organizations attempt to find in the implementation of EA, based on the claim that EA is a tool that can solve almost any type of business problem. Based on a rigorous evidence-based approach, Gong & Janssen (2019) discuss the value of and myths about enterprise architecture, recognizing that EA alone often does not provide value, but as an instrument to deal with complexity and create value.

## 3. Research method

The protocol in this research complies with the guidelines and procedures of Kitchenham (2007) for conducting Systematic Literature Reviews (SLR) in Software Engineering. It is complemented by the approach of Dybå & Dingsøyr (2008) in the sense of mapping the methodological evidence that concerns the state of the art of Enterprise Architecture application in healthcare environments. The result can help researchers understand the current leading practices, motivations that led to the choice of frameworks, methods, models, methodologies, and tools for applying EA in healthcare systems. Figure 1 shows the systematic review flow chart.

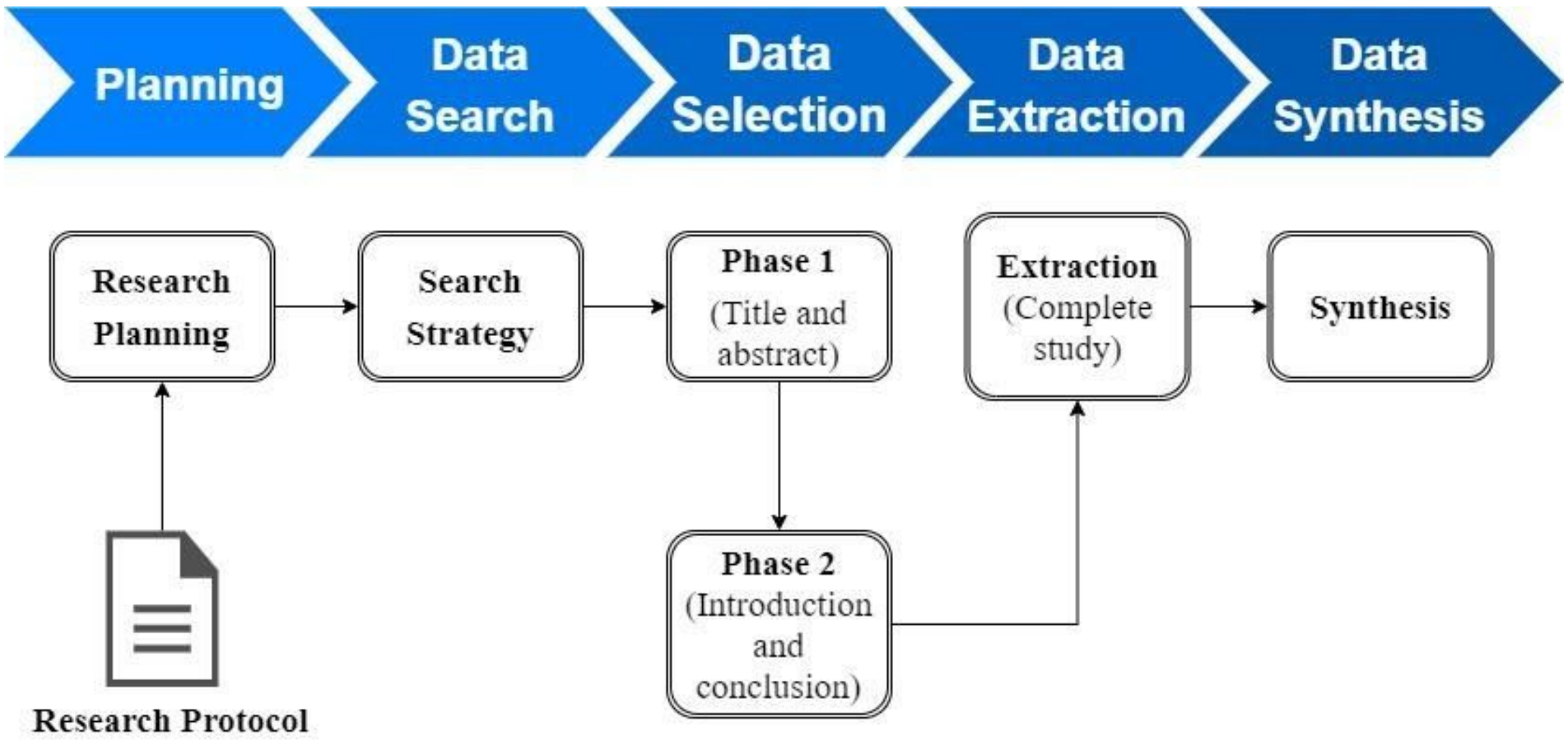

**Fig 1.** Systematic review flow chart.

### *3.1. Research Questions*

The main research question that motivates this study is: *RQ1 - What is the state of the art of the Enterprise Architecture's application in Healthcare Systems*? In other words: How has the domain of Enterprise Architecture influenced healthcare management and/or governance? Given the broad scope of RQ1, the following research questions help to map evidence that will identify specific aspects of EA's phenomenon applied in healthcare systems:

- *RQ2 - What are the most used methodologies, frameworks and best practices guide for the application of Enterprise Architecture in Healthcare systems?*
- *RQ3 - What are the most used tools and models for the development of the Enterprise Architecture in Healthcare systems?*
- *RQ4 - What are the criteria for choosing the methodology, framework, and tool used for application of the EA in Healthcare systems?*
- *RQ5 - What problems or challenges the application of EA in Healthcare systems face?*
- *RQ6 - What are the main positive impacts achieved with the application of Enterprise Architecture in Healthcare?*
- *RQ7 - What is the context of the application of Enterprise Architecture in healthcare systems?*
- *RQ8 - What are the main publication channels, and who are the most influential authors on the topic of EA in healthcare systems?*

- *RQ9 - What are the main critical success factors mentioned for the application of Enterprise Architecture in Healthcare systems?*

From the research questions above, the authors extracted the constructs shown in Table 2 to identify and codify the main characteristics found during this study.

**Table 2.** Constructs of the research questions.

| RQ | Constructs (codes) |
|---|---|
| RQ2 | 1. Methodology<br>2. Framework<br>3. Best practice Guide |
| RQ3 | 1. Tool<br>2. Model |
| RQ4 | 1. Criteria for choosing the framework/tool |
| RQ5 | 1. Problems<br>2. Challenges |
| RQ6 | 1. Positive impacts |
| RQ7 | 1. Environment/application context of EA |
| RQ8 | 1. Channels of publication<br>2. Authors |
| RQ9 | 1. Critical success factors |

### *3.2. Exclusion and Inclusion Criteria*

Table 3 shows the set of Exclusion and Inclusion criteria applied to each study of the Systematic Literature Review.

**Table 3**
Exclusion/Inclusion Criteria.

| Exclusion / Inclusion Code - (EC)/ (IC) | Description |
|---|---|
| EC01 | Studies not published in English. |
| EC02 | Studies that did not report empirical findings or literature that was only available in the form of extended abstracts, abstracts or presentations. |
| EC03 | Articles published before 2014. |
| EC04 | Secondary or tertiary studies. |
| EC05 | Studies that do not present the application of Enterprise Architecture in healthcare systems |
| EC06 | Inaccessible study. |
| IC01 | Studies that answer at least one specific research question(s). |

### *3.3. Search Strategy*

The strategy of this survey used the following scientific web databases:

- Science Direct
- IEEE Xplore Digital Library
- Scopus
- HubMed

### *3.4. Search String*

According to Kitchenham (2007), depending on the specific needs of each database search engine, SLR protocols build strings from the research question structures, and sometimes adaptations are necessary. At this point, this research string considered studies with the following terms: (1) Enterprise Architecture; (2) Health.

Terms were found anywhere in the searched documents. They were combined in boolean expression AND, adapted for each search engine, but obeying the following expression: S1 = (1) AND (2), in other words, S1= ("Enterprise Architecture") AND ("Health"). Some search engines have particularities at string expressions or limitations in their use. However, the documents collected for this survey had the relationship between Enterprise Architecture and health environments imperatively, in order to obtain results that kept the focus of this review.

### *3.5. Selection of studies*

The studies collected by the strings in the search engines went through a filtering process set in two phases. The process selected the studies according to their relevance regarding the research questions addressed in this review. In Phase 1, the protocol analyzed the studies' title, summary, and keywords, excluding the articles that could not answer any of the research questions (RQ2 to RQ9).

The articles selected in this first phase went to Phase 2, in which researchers read the studies' introduction and conclusion. In the same manner, as Phase 1, this phase eliminated studies that did not answer the research questions (RQ2 to RQ9), that is, studies that did not address the subject of this systematic review.

The selection of studies was carried out by all researchers, reducing the chances of discard relevant studies (Edwards et al., 2002). During the selection process, based on the method used by Tallon et al. (2019), the researchers worked through our entire search results to ascertain if the publications we found were relevant to a discussion of the application of EA in healthcare systems. The researchers were split into two teams, and each performed the reading and selection of all studies, according to the definitions of each phase. To assess the level of agreement between the teams, Cohen's Kappa was applied, an association measure used to describe and test the degree of agreement (reliability and precision) in the classification (Kotz et al., 2006). Landis & Koch (1977) characterized different ranges for Cohen's Kappa values, regarding the degree of agreement that these values suggest, according to the following description in Table 4:

**Table 4.** Caption for Cohen's Kappa value.

| **Value** | **Meaning** |
|---|---|
| < 0.00 | *Poor* |
| 0.00-0.20 | *Slight* |
| 0.21-0.40 | *Fair* |
| 0.41-0.60 | *Moderate* |
| 0.61-0.80 | *Substantial* |
| 0.81-1.00 | *Almost Perfect* |

The Cohen's Kappa is calculated by $Kappa = \frac{P(0) - P(E)}{1 - P(E)}$, where:

- $P(0)$: observed proportion of agreements (sum of the answers agreed divided by the total);
- $P(E)$: expected proportion of agreements (sum of the expected values of the answers agreed divided by the total).

Kappa is an interobserver agreement measure that allows for assessing if the agreement is beyond what is expected by chance, and the degree of this agreement. This measure has its maximum value as the unit value, which represents total agreement. Values close to and even below zero indicate no agreement or strong disagreements between the judges.

### *3.6. Data extraction Strategy*

After the Studies Selection phase (Phase 1 and 2), the researchers on the extraction phase read the included studies entirely (with the possibility of exclusion if there is no clear pertinence of the study to the context addressed in this survey). Data Extraction Phase seeks to answer the research questions. At this stage, all researchers independently performed the analysis and compared the results. Conflicts were resolved by consensus through a disagreement meeting.

The tool used for data extraction and synthesis was MaxQDA[1], a qualitative analysis software. In this phase, the information to be extracted from the studies were those that were related to, or that answered some specific research question. Whenever necessary, researchers took essential notes that helped in the process of synthesizing. The researchers worked separately at MaxQDA performing article extraction, following the process of joining (merge) all of the extractions using the MaxQDA itself.

### *3.7. Data Synthesis*

The adoption of this synthesis method assumes the homogeneity of the studies included in the analysis. To assist in the analysis process, we also used the MaxQDA tool in this phase to generate reports, in which it was possible to identify the correlation between the studies and the research questions, as well as to quantify these correlations with graphs and tables.

The synthesis carried out for each RQ followed specific methods adapted to each question's proposal. For RQ2, RQ3, RQ7, RQ8, and RQ9, researchers used a deductive approach, focused on the actual body of the text, in which the elements analyzed have clear and precise definitions to answer the research questions, and were classified following the explicit mention of the authors. For RQ4, RQ5, and RQ6, researchers used the method of document analysis performed by Tavakoli et al. (2017). The classifications of the coded excerpts were based on analyzes of the contextual content, using a mainly inductive approach. Due to the volume of information extracted from the questions RQ5 and RQ6, researchers create groupings of terms with semantic congruence, following the methodology of the thematic analysis coding (Ezzy, 2002) to prepare groupings of definitions and concepts found in the analysis of the extraction.

The documentary analysis allows the transition from a primary or original document to a secondary material that is an analytical and synthetic representation of the first, made through approximations that use theoretical frameworks of analysis (Bowling, 2009; Liamputtong & Ezzy, 2009). The process used tables to assist the analysis by creating semantic groups to cover the totality of the extractions of RQ5 and RQ6, being reviewed by all researchers. The divergences and additions were treated by consensus in the meetings.

## 4. Results and discussion

### *4.1. Search strategy*

The search strategy found 302 studies, according to Table 5. The search string needed to adapt to the specificities of each repository. The protocol filtered the selections by the last five years from the beginning of the research (2015-2019). To ensure the reliability of the selections, each team performed the same procedures and compared the results of the quantitative studies selected.

**Table 5.** Selection of Studies

| Search Engines (2015 - 2019) | Qty |
|---|---|
| *IEEE* | 16 |
| *SCOPUS* | 55 |
| *Science Direct* | 184 |
| *Hubmed* | 47 |
| **Total** | **302** |

### *4.2. Study Selection (Phase 1 and 2)*

In Phase 1, after the processes of eliminating duplicate studies and applying the exclusion criteria defined in the protocol, researchers analyzed the introduction, the abstract, and the keywords of the remaining 280 studies. If a study led to a divergence between the teams on the inclusion/exclusion criteria, researchers included it for Phase 2.

[1] MaxQDA - A qualitative analysis software. It can be downloaded at maxqda.com.

It resulted in a total of 68 studies, as shown in Table 6. Cohen's Kappa for Phase 1 resulted from the analyses of the two teams, which was 0.79, which represents a substantial agreement, as illustrated in Table 6.

**Table 6.** Kappa from Phase 1

| | | Team 1 | |
|---|---|---|---|
| | | Included | Excluded |
| **Team 2** | Included | 49 | 8 |
| | Excluded | 11 | 212 |
| **Total of studies selected** | | | **68** |
| **Calculated Cohen's Kappa** | | | **0,79** |

In Phase 2, the studies were gathered in PDF format, but it was not possible to access three of them integrally due to their availability; therefore, researchers excluded these studies considering exclusion criteria EC06 "inaccessible studies." These studies are: (Mocker & Ross, 2018), (Darvishzadeh et al., 2019) and (Afwani et al., 2018). Therefore, the introductions and conclusions of the 65 studies were analyzed simultaneously with eleven exclusions. Cohen's Kappa was also used in this phase and was scored 0.59 (moderate agreement level), according to Table 7.

**Table 7.** Kappa from Phase 2

| | | Team 1 | |
|---|---|---|---|
| | | Included | Excluded |
| **Team 2** | Included | 44 | 1 |
| | Excluded | 9 | 11 |
| **Total of studies selected (after meeting)** | | | **49** |
| **Calculated Cohen's Kappa** | | | **0,59** |

Following the methodology proposed in the protocol, the two teams resolved the disagreements resulting from Phase 2 through a "disagreement meeting." They reconsidered the disagreements by re-reading the introduction and conclusion of the ten studies in question. By consensus, they decided to include 5 of them for Phase 3, resulting in a total of 49 studies.

### *4.3. Data extraction (Phase 3)*

In phase 3, each team read the 49 studies in full, in which there was a consensus to exclude three more articles that did not answer any of the research questions in this systematic review, totaling 46 studies. The researchers used MaxQDA to conduct the entire extraction process, a qualitative analysis tool used to categorize relevant information through the use of codes, colors, symbols, or even emoticons. They perform statistical analysis of these data, allowing a holistic view of all work done on the software.

The segment encodings and annotations made in the studies using MaxQDA were exported, through the software itself, in a spreadsheet in .xls format, and used for team analysis in the data synthesis phase.

### *4.4. Synthesis (Phase 4)*

In phase 4, the researchers conducted the data synthesis process on a thorough analysis of the spreadsheets and graphs resulting from the extraction process carried out by the two teams of researchers using the MaxQDA tool and Excel. The two teams analyzed the coded excerpt and annotations in the 46 studies to see if there was any inconsistency in the relationship between the extracted segments and the research questions. Figure 2 shows the graph with the distribution of these studies in the years 2015 to 2019, which represents an average of 9 studies published per year.

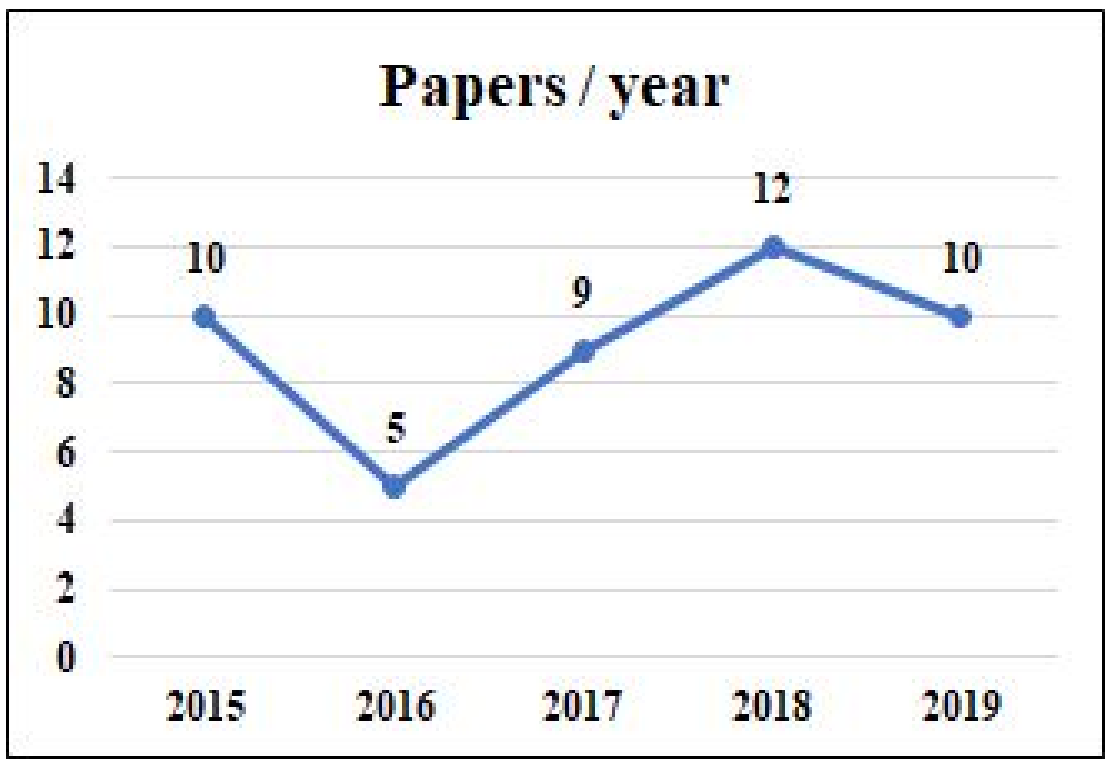

**Fig 2**. Number of studies per year

In Figure 3, the word cloud, generated from the 46 studies, is illustrated, in which approximately 15500 words are present. The most common words, excluding connectors such as “the,” “of,” “in,” that do not add value to the formation of the cloud were “architecture”, “health”, “information”, “enterprise”, “business”, “data” and “healthcare”.

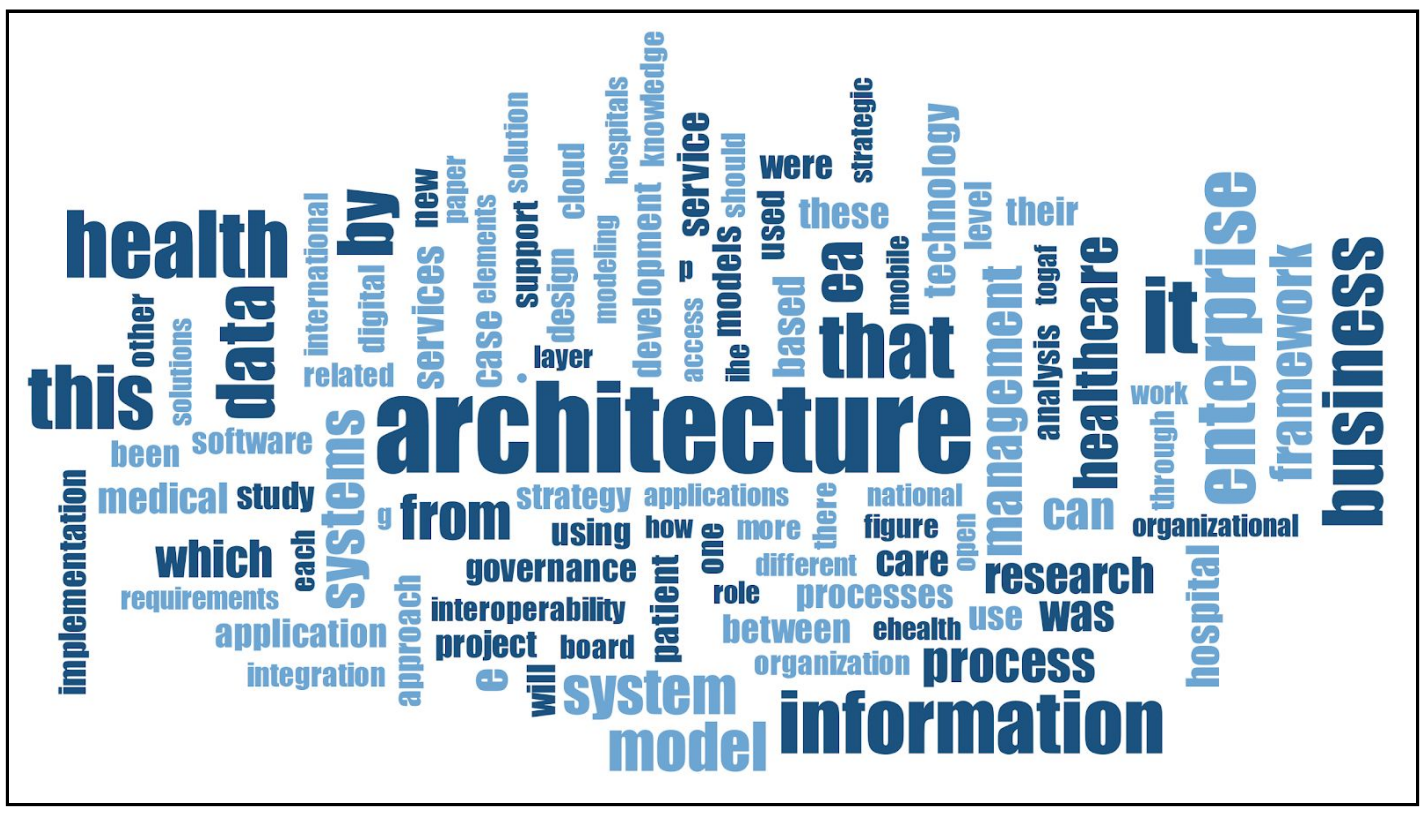

**Fig 3.** Word cloud of the 46 studies

To simplify the arrangement of data in the tables that respond to the RQ’s, we created equivalence codes for each of the 46 references of the selected studies, as shown in Table 8. Sections 4.4.1. to 4.4.9. presents the research questions, and the results found by the teams in the data extraction process.

**Table 8**
Equivalence code for references

| Author | Cod. | Author | Cod. |
|---|---|---|---|
| (Nugraha et al., 2017) | A1 | (Olsen, 2017) | A24 |
| (Yamamoto & Traoré, 2017) | A2 | (Eldein et al., 2017) | A25 |
| (Memon et al., 2019) | A3 | (Ateetanan et al., 2017) | A26 |
| (Wautelet, 2019) | A4 | (Bygstad & Hanseth, 2016) | A27 |
| (Zwienen et al., 2019) | A5 | (Handayani et al., 2019) | A28 |
| (Mayakul et al., 2018) | A6 | (Traoré & Yamamoto, 2018) | A29 |
| (Masuda et al., 2019b) | A7 | (Haghighathoseini et al., 2018) | A30 |
| (Feltus et al., 2015) | A8 | (Bakar & Selamat 2016) | A31 |
| (Mayakul & Kiattisin, 2018) | A9 | (Javed et al., 2015) | A32 |
| (Motoc, 2017) | A10 | (Fossland & Krogstie, 2015) | A33 |
| (Tarenskeen et al., 2015) | A11 | (Stäubert et al., 2015) | A34 |
| (Yamamoto & Zhi, 2019) | A12 | (Pankowska, 2015) | A35 |
| (Lessard et al., 2017) | A13 | (Vinci et al., 2016) | A36 |
| (Masuda et al., 2018) | A14 | (Tarenskeen et al., 2018) | A37 |
| (Purnawan & Surendro, 2016) | A15 | (Pankowska, 2018) | A38 |
| (Masuda et al., 2019a) | A16 | (Masuda et al., 2017) | A39 |
| (Gebre-Mariam & Fruijtier, 2018) | A17 | (Winter et al., 2018) | A40 |
| (Ahmad et al., 2018) | A18 | (Beštek & Stanimirović, 2017) | A41 |
| (Mousavi et al., 2018) | A19 | (Kaushik & Raman, 2015) | A42 |
| (Rijo et al., 2015) | A20 | (Gebre-Mariam & Bygstad, 2016) | A43 |
| (Adenuga et al., 2015) | A21 | (Herdiana, 2018) | A44 |
| (Yamamoto et al., 2019) | A22 | (Noran, 2015) | A45 |
| (Ajera et al., 2019) | A23 | (Toma et al., 2019) | A46 |

#### *4.4.1. RQ2 – What are the most used methodologies, frameworks and best practices guide for the application of Enterprise Architecture in Healthcare systems*

To answer the RQ2, the researchers considered data extracted from the studies referring to which methodologies cover the development of EA in healthcare systems. Table 9 shows that the TOGAF framework was applied in 11 studies, representing 22% of the applications, followed by AIDAF, with 5 (11%), Weil, and Ross with 3 (6%) and Zachman's framework, applied in 2 studies (4%). These four frameworks were the most used in Health EA applications, representing a total of 43%. Many of these 46 selected studies performed combinations between frameworks, methods, methodologies, or proper practice guides to achieve a broader scope of work. In A30, the authors apply TOGAF alongside Kendall's W method (Okoli & Pawlowski, 2004), a non-parametric statistical method used to assess agreement between evaluators. They used AHP (Vargas, 2010), a method to support decision-making, and ANOVA (Statistics, 2018), a one-way variance analysis method used to determine whether there are statistically significant differences between the means of three or more independent (unrelated) groups. In Table 9, it is also possible to verify the countries where these frameworks were applied, the total number of studies and their references.

**Table 9**
Most used methodologies/frameworks/best practices

| Methodology /Framework / Best practices guide | Country | Nº Papers | Identified studies |
|---|---|---|---|
| TOGAF (The Open Group Architecture Framework) | Indonesia, Japan, Iran, Malaysia, Netherlands, Poland | 11 | A1, A2, A10, A25, A28, A29, A30, A31, A37, A38, A44 |
| AIDAF (Adaptive Integrated Digital Architecture Framework) | Japan, Germany | 5 | A7, A14, A16, A39, A46 |
| Weil and Ross | Netherlands, India, Norway | 3 | A11, A42, A43 |
| CVI (Content Validity Index ) | Thailand, Iran | 2 | A6, A19 |
| Delphi Technique/method | Thailand, Iran | 2 | A9, A30 |
| Zachman's framework | South Africa, Brazil | 2 | A21, A36 |
| IDEFØ | Norway, Australia | 2 | A33, A45 |
| O-DA (Open Dependability through Assuredness) | Japan | 1 | A2 |
| ISO 42030 - Architecture Evaluation Framework | Australia | 1 | A3 |
| I-Tropos | Belgium | 1 | A4 |
| MoDrIGo standing for Model-Driven IT Governance | Belgium | 1 | A4 |
| NFR tree | Belgium | 1 | A4 |
| Design Science | Netherlands | 1 | A5 |
| i* framework | Thailand | 1 | A6 |
| JCI (Joint Commission International) | Thailand | 1 | A6 |
| Kappa | Thailand | 1 | A6 |
| ADR (Action Design Research) | Belgium | 1 | A8 |
| BIE Generic Schema | Belgium | 1 | A8 |
| An e-health Enterprise Architecture framework | Thailand | 1 | A9 |
| DM (Design Matrix) | Netherlands | 1 | A11 |
| FEAF (Federal Enterprise Architecture Framework) | Canada | 1 | A13 |
| ESIA Method | Indonesia | 1 | A15 |
| ANT (Actor–Network Theory) | Ethiopia | 1 | A17 |
| BPAOntoEIA framework | Jordan | 1 | A18 |
| Riva Method | Jordan | 1 | A18 |
| Expert Panel Method | Iran | 1 | A19 |
| ISO TR 14639 | Iran | 1 | A19 |
| Gartner | Portugal | 1 | A20 |
| MBJT (Model Based Jobs Theory) | Japan | 1 | A22 |
| FAIR (Factor Analysis of Information Risk) | Japan | 1 | A29 |
| AHP (Analytical Hierarchy Process) | Iran | 1 | A30 |
| ANOVA (One-way Analysis Of Variance) | Iran | 1 | A30 |
| Kendall's W | Iran | 1 | A30 |
| BSC (Balanced Scorecard) | Malaysia | 1 | A31 |
| Malaysian Public Sector Enterprise Architecture Framework - 1 Government Enterprise Architecture (1GovEA) | Malaysia | 1 | A31 |
| Enterprise architecture planning (EAP) | Germany | 1 | A34 |
| FHIR (Fast Healthcare Interoperability Resources) | Germany | 1 | A40 |
| HL7 Clinical Document Architecture | Germany | 1 | A40 |
| ISO Standard 14721:2012 | Germany | 1 | A40 |
| Continua Health Alliance | Slovenia | 1 | A41 |
| IHE (Integrating the Healthcare Enterprise) | Slovenia | 1 | A41 |
| OpenEHR | Slovenia | 1 | A41 |
| SNOMED (Systematized Nomenclature of Medicine) | Slovenia | 1 | A41 |
| GERAM - ISO (Generalised Enterprise Architecture and Methodology) | Australia | 1 | A45 |

#### *4.4.2. RQ3 – What are the most used tools and models for the development of the Enterprise Architecture in Healthcare systems?*

In Table 10, there are tools and models used in the development of EA in Healthcare environments, corresponding to RQ3. The most applied software was Archimate EA; it was found in nine studies, representing 17% of the applications, followed by BPMN with 6% and 3LGM² with 4%. Some studies combine different tools and models in a unique Enterprise Architecture strategy. As can be seen in A38, this paper applies the Archimate tool, BPMN, and CMMN models to perform the modeling of EA for hospitals.

**Table 10**
Most used tools/models

| Tool / Model | Country | Nº Papers | Identified studies |
|---|---|---|---|
| ArchiMate EA | Indonesia, Belgium, Japan, Poland | 9 | A1, A2, A8, A12, A22, A25, A29, A35, A38 |
| BPMN (Business Process Model and Notation) | Thailand, Norway, Poland | 3 | A26, A33, A38 |
| Likert scales | Thailand, Iran | 2 | A9, A19 |
| Ampersand | Netherlands | 2 | A11, A37 |
| 3LGM² (Three Layer Graph Based Meta Model) | Germany | 2 | A34, A40 |
| CASE (DesCARTES Architect) | Belgium | 1 | A4 |
| DyAMM | Netherlands | 1 | A5 |
| ZiRA | Netherlands | 1 | A5 |
| SWOT | Thailand | 1 | A6 |
| Node-RED | Thailand | 1 | A7 |
| Ptolemy | Thailand | 1 | A7 |
| ReMMo (Responsibility metamodel) | Belgium | 1 | A8 |
| RBAC | Belgium | 1 | A8 |
| Reference domain model for Hospitals | Netherlands | 1 | A11 |
| V Model | Netherlands | 1 | A11 |
| Consolidated Reference Model (by FEAF) | Canada | 1 | A13 |
| BTEP (Business Transformation Enablement Program) | Indonesia | 1 | A15 |
| Java-based OWL APIs | Jordan | 1 | A18 |
| Limesurvey | Iran | 1 | A19 |
| The Essential Project EA tool | Portugal | 1 | A20 |
| Troux EA tool | Norway | 1 | A24 |
| Service Blueprint (SB) | Thailand | 1 | A26 |
| CPM (Configurable Process Model) | Malaysia | 1 | A31 |
| SOA (Service-Oriented Architecture) | Pakistan | 1 | A32 |
| CMMN | Poland | 1 | A38 |
| STRMM (STrategic Risk Mitigation Model) | Japan | 1 | A39 |
| OAIS (Open Archival Information System) | Germany | 1 | A40 |

#### *4.4.3 RQ4. – What are the criteria for choosing the methodology, framework, and tool used for application of the EA in Healthcare systems?*

In order to answer the RQ4, the researchers faced particular difficulty in the extraction of the excerpts because not all selected studies demonstrated which criteria they used to choose the framework, methodology, or tool to develop the Enterprise Architecture in the Healthcare system. Since it is a subjective question, it is necessary to deepen further the practical results achieved by these selected studies to answer this specific research question.

Eleven studies chose TOGAF as Enterprise Architecture methodology with different choice criteria, but some do not explicitly state it. In Nugraha et al. (2017), TOGAF is the selected framework because it has TOGAF ADM with several phases that facilitate the construction of enterprise architecture. The method is detailed, flexible, and adjustable according to changes and demands of engineering, in addition to being open-source. According to Eldein et al. (2017), TOGAF describes required business and ICT architecture. Also, it provides a step by step approach in building and implementing enterprise architecture.

Handayani et al. (2019) developed a corporate architecture (EA) for a health referral information system (HRIS), including individual healthcare in Indonesia. They decided to choose TOGAF based on empirical and exploratory studies conducted in healthcare organizations. Tarenskeen et al. (2018) decided on the application of TOGAF because it is relevant for matching existing applications to a Radical Business Requirements Change. It serves as a roadmap for the transformation of a Base Architecture (AS-IS) to a Target Architecture (TO-BE). Herdiana (2018) concluded that TOGAF could be used to develop a wide range of enterprise architecture in conjunction with any other framework that focuses on a particular sector as designed as a generic framework.

Yamamoto & Traoré (2017) propose the O-DA (Open Dependability through Assuredness) standard, which applies in a case study on the African Healthcare Information System. O-DA was used to mitigate risks, for modeling dependencies, building assurance cases, and achieving agreement on accountability on the complex interoperable systems. Memon et al. (2019) recognized that ISO 42030 contributes to the maturity of architecture governance because it systematizes the elements to be considered by a process that supports architectural decision making.

Wautelet (2019) developed a framework called MoDrIGo, standing for Model-Driven IT Governance. It considers business IT services in as-is and to-be specifications to specifically support governance decisions, as well as, is made to perform at best in pure organizational i* modeling. Mayakul et al. (2018) justified the i* methodology is suitable to help them understand the primary resources and information flow within the enterprise at an early stage. The i* can present the relationship between entities and the contribution to the visibility of the information. At the same time, they used the international standard and quality control JCI because that is a global gold standard to perform as a standard regulator, advising and facilitating a healthcare organization towards better performance and outcome.

Masuda et al. (2018) and Masuda et al. (2017) choose the Adaptive Integrated Digital Architecture Framework (AIDAF) based on adaptive enterprise service system logic expanding on the system of systems (SoS) and agility. At the same time, Toma et al. (2019) consider AIDAF an adaptive EA cycle that makes provisions for project plan and architecture design documents for new Digital IT related projects drawn up on a short-term basis. In addition to the fact that AIDAF is capable of flexibly adapting to new Digital IT projects continuously. Mayakul & Kiattisin (2018) use the method to information systems research, called technique Delphi, which has benefits for planning, needs assessment, policy determination, and resource utilization.

Lessard et al. (2017) present an architecture framework for LHS (Learning Health Systems), based on the Federal Enterprise Architecture Framework (FEAF) because the FEAF captures an organization's or system's human and technical components, enabling the alignment of multi-stakeholder goals within an organization's structure and technical systems. According to the author, the FEAF provides an ideal basis for LHS architectures situated in multi-professional health systems, such as hospitals or health maintenance organizations.

Ahmad et al. (2018) state that Enterprise Architecture methods lack the knowledge of business processes in an enterprise. Therefore, the authors applied the BPAOntoEIA framework, which provides a semi-automatic semantic derivation of information categories from the Riva-based business process architecture of an organization. Rijo et al. (2015) decided, according to the goals for a proof of concept, to follow the aspects of the Gartner pragmatic approach, which is to create a shared vision among business owners, information specialists, and the technology implementer to drive profitability.

Adenuga et al. (2015) propose an Enterprise Architecture solution considering integration and interoperability issues while Vinci et al. (2016) describe an evaluation model to a regional health network management, both use in its solutions the Zachman's framework. The first justified the choice because the framework helps managers communicate efficiently and map enterprise architecture as a foundation for discussion that facilitates change. The second study considers Zachman's framework most suitable due to its clarity and objectivity to acquire information in a healthcare system.

Yamamoto et al. (2019) use the Model-Based Jobs Theory (MBJT) because it fosters consistent visual modeling methods and integrates innovation and enterprise architecture using the ArchiMate tool, in addition to easily integrating MBJT and ConOps. Traoré & Yamamoto (2018) applied Factor Analysis of Information Risk (FAIR) methodology because it helps enterprises communicate on their risk information, thus aligning with the enterprise's needs through risk scenarios analysis and assessment analysis. Fossland & Krogstie (2015) adopted a top-down generic model IDEF0 since it is the best practice for logical/generic/conceptual process models.

Stäubert et al. (2015) adopted enterprise architecture planning (EAP) because it is a method capable of designing or changing an information system according to the strategic goals of an enterprise. They also chose 3LGM2 because element types or using wildcards in the name or description fields enables the user to find IHE (Integrating the healthcare enterprise) concepts and because Enterprise architecture planning (EAP) tools like the $3LGM^2$ tool help build up and analyze Information System models.

Beštek & Stanimirović (2017) applied the openEHR tool and the systematized terminology of Medicine SNOMED to define clinical data used for exchange over Integration Health Enterprise (IHE). OpenEHR tooling supports the modeling of core artifacts that are publicly available and consider SNOMED as the central terminology for mapping other existing terminologies because it is an ontology that enables complex relationships between the terms. They also adopted the guidelines of Continua Health Alliance in combination with IHE to exchange data between Electronic Health Record (EHR) and Personal Health Record (PHR) in a more suitable way, despite identified gaps and limitations.

Lessard et al. (2017) analyzed that BSC is a method that helps identify the most important goals for an organization's performance and then enables the organization to monitor their achievement and impact on one another through a set of measures. Ahmad et al. (2018) adopt the Riva BPA design, an object-based approach with its foundation in the classical business analysis phase of the information engineering paradigm. They consider Riva BPA indicated for enterprise business process architecture.

Archimate was the most used model in the selected studies. Nugraha et al. (2017) selected Archimate to define a model to describe the development and operation of the business process, organization structure, and information path. It is a modeling standard language for enterprise architecture, and it is distinguished for its openness and independence. Its specification helps many enterprise architects explain, analyze, and visualize the relationships across business domains in less ambiguous ways. Furthermore, it can model general enterprise architecture in different areas.

Traoré & Yamamoto (2018) emphasize that ArchiMate is an Enterprise Architecture visual language with a set of default iconography for describing, analyzing, and communicating many EA's concerns as they change over time. According to Pankowska (2015), the ArchiMate Canvas Model allows us to catch intangible requirements and emphasize the stakeholders' place in the system architecture. Pankowska (2018) chose it because its language and software tools are the most suitable for strategic issues visualization and analysis. Zwienen et al. (2019) adopt DyAMM as it is an existing model to serve as a basis for tailoring and also because ZiRA incorporates the DyAMM. They consider that the ZiRA components are mostly product-oriented.

Mayakul et al. (2018) chose to use SWOT analysis, considering it is a standard analytical tool for strategic planning and policy implementation in various businesses. The BTEP was used in Purnawan & Surendro (2016) as a preferred methodology by TOGAF to assess business transformation readiness. Mousavi et al. (2018) chose Limesurvey because it is an online open-source tool for conducting a survey and performing the analysis. In Rijo et al. (2015), the choice was for the "The Essential Project tool," instead of ArchiMate, because the alignment between ArchiMate and TOGAF, making the use of this walkthrough more difficult, once of the framework used in this work was that of Gartner. The Essential Project was also chosen because it is open source.

It has found selection criteria for BPMN in three papers. Ahmad et al. (2018) mentioned that Business process models of the enterprise enrich semantically using the instantiated BPMN 2.0. Ateetanan et al. (2017) described that BPMN is a business process modeling standard and, indeed, the most used language for diagrammatically representing processes. It provides a standard business process model notation for describing and analyzing the business process in detail. Pankowska (2018) emphasizes that BPMN is dominant for business analytics, assuming BPMN can support business process orientation, as a more detailed analysis of researchers' tasks. This paper also used the CMMN, which reported that CMMN modeling provides some essential values to the business architecture modeling. Sometimes, in the domain of business process, modeling a certain degree of flexibility is required.

Winter et al. (2018) used $3LGM^2$ for modeling health information systems, especially trans-institutional information systems, and, therefore, the entire information system of SMITH (Smart Medical Information Technology for Healthcare). Winter et al. (2018) also used OAIS and justified that this model provides a framework, including terminology and concepts, to describe and compare architectures and operations of archives. Thus, for sharing their content, OAIS is the most common standard for archival organizations (ISO Standard 14721:2012).

Masuda et al. (2017) applied the STRMM (STrategic Risk Mitigation Model) model as the Risk Mitigation model in the Architecture Board. It is based on the case study research Masuda et al. (2018) that verify that the Architecture Board can control the Solutions with "STRMM model for Digital Transformation."

#### *4.4.4. RQ5 – What problems or challenges the application of EA in Healthcare systems face?*

In response to RQ5, the researchers listed the main problems and challenges found in the selected studies related to the application of EA in Healthcare Systems, and grouped them into macro-categories according to the context of the issues, as shown in Table 11. They conducted the categorization through semantic congruence between the extracted excerpts. For instance, in A3, the author describes that "the health enterprise is a complex evolving system of systems (SoS) both on national and global scales." In A4, they suggest the administrative activities became more and more complicated. Therefore, given the semantic congruence of these segments, the category "organizational/cultural complexity of health environments" was created.

**Table 11**
Main problems and challenges in implementing EA

| Problem/Challenge Category | Nº Studies | Study Reference |
|---|---|---|
| Organizational complexity of health environments | 13 | A3, A4, A10, A12, A15, A20, A23, A24, A26, A28, A31, A34, A43 |
| Difficulty in integrating/accessing data of various kinds | 8 | A2, A6, A9, A12, A14, A15, A20, A32 |
| Heterogeneous stakeholder interests; Communication problems | 7 | A12, A14, A15, A20, A24, A27, A31 |
| There is no clear definition of the organization's objectives/goals/processes; lack of organizational maturity | 7 | A15, A20, A21, A24, A26, A27, A43 |
| Privacy and data security | 6 | A3, A14, A17, A21, A23, A29 |
| Lack of an appropriate model/aligned to the needs/infrastructure of the organization | 6 | A1, A8, A13, A14, A24, A32 |
| Organizational / IT capacity | 4 | A15, A17, A20, A31 |
| Lack of skilled professionals | 4 | A3, A17, A21, A31 |
| Political instability, constant organizational changes, laws, rules | 4 | A3, A6, A24, A31 |
| Costs | 3 | A15, A20, A21 |

It is possible to conclude that the four biggest problems/challenges encountered in the application of EA in healthcare systems are: (i) organizational/cultural complexity of each health environments; (ii) difficulty in integrating or accessing data of various kinds; (iii) varied stakeholder interests and communication problems; and (iv) lack of clear definition of the organization's objectives/goals/processes; lack of organizational maturity.

#### *4.4.5. RQ6 – What are the main positive impacts achieved with the application of Enterprise Architecture in Healthcare systems?*

RQ6 sought to capture the main positive impacts achieved with the application of EA in Healthcare systems. In Table 12, the authors extract 68 excerpts from 28 studies that clearly explained findings for this research question. As in Table 11, the authors grouped the categories that had semantic congruence into categories, listed in order of the most cited positive impacts. Among these, the three most reported are: "Describes and categorizes the architecture and operation of business processes, organizational structure, and data to facilitate the acquiring information," mentioned in 20% of studies. Second, "it benefits from change management, process and quality improvement," present in five reviews (11%), followed by "systematizes the elements to be considered for decision making," in 5 (11%).

**Table 12**
Positive Impacts

| Positive Impacts | Nº studies | Reference Studies |
|---|---|---|
| Describes and categorizes the architecture and operation of business processes, organizational structure, and data to facilitate the acquiring information | 9 | A1, A2, A4, A13, A19, A20, A28, A31, A42 |
| Benefits in change management, process and quality improvement | 5 | A13, A18, A19, A36, A43 |
| Systematizes the elements to be considered for decision making | 5 | A3, A13, A20, A42, A43 |
| Contributes to the maturity of management and governance | 4 | A3, A5, A13, A20 |
| Link business strategy, business operations and IT | 3 | A2, A4, A36 |
| Assists in the development and management of projects and processes | 3 | A8, A13, A28 |
| Offers greater consistency and comprehensibility | 3 | A2, A10, A20 |
| Improves alignment between standards, security controls and legislative privacy measures | 2 | A2, A13 |
| Contributes to cost reduction | 1 | A1 |
| Facilitates the revolution and application of technology system | 1 | A1 |
| Contributes to problem management | 1 | A2 |
| Determines new organizational needs | 1 | A2 |
| Assists in the alignment and identification of goals and objectives | 1 | A13 |
| Allows simulation of possible business strategies as problem-solving | 1 | A20 |
| Collect lessons learned | 1 | A20 |
| Enables better alignment between stakeholders | 1 | A33 |

#### *4.4.6. RQ7 – What is the context for the application of Enterprise Architecture in healthcare systems?*

RQ7 sought information about the Healthcare environment or context of Enterprise Architecture application. In Table 13, it presents that the most significant application was found in hospitals, followed by implementations of EA in digital health (e-Health) and Health Information System. The number of studies carried out in hospitals adds up to a total of 14 (30%); for applications in e-Health, 11 studies (24%), and 8 (17%) studies in SIS.

**Table 13.** EA Environment /Application Context

| EA Environment/Application Context | No. studies | Reference Studies |
|---|---|---|
| Hospital | 14 | A4, A5, A6, A8, A11, A15, A16, A18, A20, A23, A26, A30, A37, A38 |
| Digital health (e-Health) | 11 | A10, A12, A19, A21, A22, A25, A27, A29, A35, A40, A46 |
| Health Information System (HIS) | 8 | A2, A3, A13, A17, A28, A41, A43, A45 |
| Public health system | 7 | A9, A24, A31, A33, A36, A42, A44 |
| Healthcare community (pharmaceutical companies, healthcare companies, etc.) | 5 | A7, A14, A32, A34, A39 |
| Primary Health Care Unit | 1 | A1 |

Some studies did not identify the organizational type of the hospital where the research was carried out, public, private, or university. Among the studies that provided information, three were in public hospitals, such as A8, A11, and A26; two in private hospitals, A6 and A15; and one in a university hospital, A30.

### *4.4.7. RQ8 – Who are the main publication channels and the most influential authors on the topic of EA in Healthcare systems?*

In response to RQ8, the researchers listed the fifteen main publication channels, listed in Table 14. The order established was for the channels that had more publications on the theme proposed in this review, followed by relevance, considering their impact factor. 37% of publications on the topic were made by the first five publication channels, with emphasis on Smart Innovation, Systems, and Technologies, with five studies published on the topic, representing 11% of the total.

**Table 14**
Publication Channels

| Conference / Journal | Qty. of publications | % | Impact Factor |
|---|---|---|---|
| Smart Innovation, Systems and Technologies | 5 | 11% | 0.59 |
| IIAI International Congress on Advanced Applied Informatics (IIAI-AAI) | 4 | 9% | 0.42 |
| Procedia Computer Science | 4 | 9% | 1.26 |
| CEUR Workshop Proceedings | 2 | 4% | 0.34 |
| Studies in Health Technology and Informatics | 2 | 4% | 0.44 |
| Government Information Quarterly | 1 | 2% | 6.43 |
| Journal of Systems and Software | 1 | 2% | 4.02 |
| International Journal of Medical Informatics | 1 | 2% | 3.02 |
| Healthcare Informatics Research | 1 | 2% | 2.87 |
| Procedia CIRP | 1 | 2% | 2.10 |
| International Conference on Research Challenges in Information Science (RCIS) | 1 | 2% | 1.02 |
| Methods of Information in Medicine | 1 | 2% | 1.11 |
| European Conference on Information Systems, ECIS | 1 | 2% | 1.05 |
| International Journal of Biomedical Engineering and Technology | 1 | 2% | 0.57 |
| International Journal of Enterprise Information Systems | 1 | 2% | 0.71 |
| Others | 19 | 41% | - |
| Total | 46 | 100% | - |

Also, in response to RQ8, 144 authors who published the 46 selected studies were found. Table 15 lists the authors who published two or more articles from this selection, including the area of application of Enterprise Architecture in Healthcare Systems in which their publications addressed. As can be seen, nine studies (20%) were published by the three principal authors, addressing Enterprise Architecture in the Health Information System, e-Health, Hospital, and Community of Health. These authors had some publications together. Authors Seiko Shirasaka and Yoshimasa Masuda were also present in some publications by Shuichiro Yamamoto.

Dr. Shuichiro Yamamoto is currently a professor at the Graduate School of Informatics at Nagoya University. His current research includes Digital Balanced Scorecard toward Digital Transformation, DX Visualization Approach Using ArchiMate, and Tailoring Approach on Enterprise Architecture Framework towards DX. Dr. Yoshimasa Masuda currently works at the Computer Science Department, Carnegie Mellon University. Their current project is 'Digital architecture framework,' and MSc. Seiko Shirasaka is a professor of the Graduate School of System Design and Management (SDM), Keio University. His fields of specialty include systems engineering, innovation, innovative design, concept engineering, model-based development, space systems engineering, system assurance, functional safety management, and standardization.

**Table 15**
Main authors

| Main authors | Nº papers | Papers | EA application context |
|---|---|---|---|
| Shuichiro Yamamoto | 9 | A2, A7, A12, A14, A16, A22, A29, A39, A46 | Community of Health, e-Health, Hospital, SIS |
| Yoshimasa Masuda | 4 | A7, A14, A16, A46 | Community of Health, e-Health, Hospital, SIS |
| Seiko Shirasaka | 3 | A7, A14, A39 | Community of Health |
| Ovidiu Noran | 2 | A3, A45 | SIS |
| Tetsuya Toma | 2 | A16, A46 | e-Health, Hospital |
| Thomas Hardjono | 2 | A14, A39 | Community of Health |
| Malgorzata Pankowska | 2 | A35, A38 | e-Health, Hospital |
| Mariam Traoré | 2 | A2, A29 | e-Health, SIS |
| Rui Pedro Charters Lopes Rijo | 2 | A20, A36 | Hospital, Public Health System |

#### *4.4.8 RQ9 – What are the main critical success factors mentioned for the application of Enterprise Architecture in Healthcare systems?*

The RQ9 aimed to capture the critical success factors reported by the authors in the implementation of Enterprise Architecture in Healthcare systems. Among the 46 studies analyzed, only two mentioned eight factors listed in Table 16. Other studies have had successful cases in the implementation of EA in Healthcare. However, the authors did not demonstrate the critical success factors, even the researchers considering lexical research supported by MaxQDA tool to reinforce the term's capture.

**Table 16**
Critical Success Factors

| | Critical success factors | | Papers |
|---|---|---|---|
| - | Commitment from CIO and top management | | A14 |
| - | Collaboration between the architecture and PMO communities on Digital platforms | | |
| - | Internal Process Perspective | 1. business driven approach; 2. clear communication; 3. strong governance; 4. mutual understanding; 5. clear planning, scope and coverage; 6. standard rules and EA process | A31 |
| - | Learning and Growth Perspective | 1. systematic assessment mechanism; 2. complete documentation; 3. learning culture; 4. skillful architect; 5. relevant training and certification. | |
| - | Authority Support Perspective | 1. continuous support; 2. EA recognition; 3. mandated EA rules and processes; 4. positive political influence; 5. stakeholder participation | |
| - | Cost Perspective | 1. enough resources financial allocated; 2. economic pressure; 3. enough supply of other resources | |
| - | Technology Perspective | 1. Easy to use EA tools; 2. Standard tools, methodology, EA model or artefact | |
| - | Talent Management | 1. Retention of expertise | |

The critical success factors mentioned in A14 were the application of EA in the global healthcare enterprises (GHE) for solving issues and mitigating related architecture risks while implementing AIDAF. The authors formulated the useful elements of risk mitigation strategy with the Architecture Board and clarify the challenges and critical success factors of architecture reviews on digital platforms in the Architecture Board for EA practitioners.

A31 described the experience of implementing EA in the public sector, in which a case study was carried out at the Malaysian Ministry of Health (MOHM) and identified six categories of critical success factors that allowed this implementation to be successful, and which, according to the author, can be guidelines for other public organizations.

## 5. Conclusion

Enterprise Architecture is currently present in several business branches, and scientific literature discusses it widely, with professionals and researchers studying and applying its concepts throughout the world. There is a diversity of methodologies, tools, and frameworks available, justifying the large number of diverse organizations that have used EA for management support and applied governance. Choosing what methodologies or tools are most appropriate could be costly and a complicated task. There are no standardized guidelines to implement Enterprise Architecture in a specific field (Purnawan & Surendro, 2016), which requires the ability to provide adaptations that meet the requirements of each company.

Most publications concerning the implementation of EA lack detailed and accurate information about the application environment or other data that could be relevant for the dissemination of good practices and the success achieved. Their inaccuracy and lack of detail often make the extraction of data a challenging job for conducting research. For instance, in some studies of this review, there was a lack of detail related to the characterization of the hospital where the research was conducted, leaving some question marks such as "is it a small, medium or large hospital?", or "is it a clinical or emergency hospital?".

On the other hand, studies that have provided details about the nature of the health environment offered a significant scientific basis for other organizations that seek methodologies, methods, and tools to assist their management and governance. These studies can become an essential empirical basis in selecting a set of good practices and making it possible to carry out new relevant studies. Given the difficulties encountered concerning the detailed level of some studies, the researchers inferred that it occurs due to the insecurity of sharing certain types of data from companies or because of ethical or cultural reasons.

The result of this study elucidates how researchers and professionals in the fieldwork with Enterprise Architecture applied the concepts and practices to healthcare systems, in addition to some criteria used for their choice. We also selected the main positive impacts that were described by the authors, based on results achieved by an empirical approach, including critical success factors in some of these applications. Besides, this work brings the main publication channels and the most influential authors on the topic of EA in Healthcare Systems.

This paper's primary motivation was to fill the gaps found in the current literature of systematic reviews and systematic mappings concerning success cases in the application of EA. Thus, this work described the state of the art related to the application of Enterprise Architecture in Healthcare Systems, focusing on specific research questions that have made it possible to reveal practical aspects of EA implementation. By answering these research questions, this SLR contributes as a repository of relevant data to help researchers find successful EA cases in the healthcare environment and understand its implementation.

The data collected can, therefore, help researchers obtain information that will support them in spreading knowledge about EA, encouraging the production of new scientific and practical work in the field. Although we have a clearly defined scope of our work, the subject addressed is quite broad. It may stimulate the development of several other specific research questions that would further explain this phenomenon. Hence, we expect that this study will be a driving factor for researchers to conduct new SLRs and expand the understanding of the phenomenon of the application of Enterprise Architecture in healthcare systems.